# INNOVATING THE SOFTWARE ENGINEERING CLASS THROUGH MULTI-TEAM DEVELOPMENT

A. Brockenbrough

*Salem State University (UNITED STATES)*

## Abstract

College computer science courses do well at educating students on how to design and implement solutions to simple problems. In beginning courses, the student is challenged to solve simple problems, limited to what can be coded by a single student in a few days or weeks. After the students have mastered the basic skills, the curriculum expands to more complex topics including software engineering. A class in software engineering introduces students to a systematic, defined, and quantifiable approach to the development of software. By using a systematic approach, more complex problems can be tackled, usually involving multiple developers.

Often software engineering classes have the student concentrate on designing and planning the project but stop short of actual student team development of code. This leads to criticism by employers of new graduates that they are missing skills in working in teams and coordinating multiple overlapping changes to a code base. Additionally, students that are not actively experiencing team development are unprepared to understand and modify existing legacy-code bases written by others.

This paper presents a new approach to teaching undergraduate software engineering that emphasizes not only software engineering methodology but also experiencing development as a member of a team and modifying a legacy code base. Our innovative software engineering course begins with learning the fundamentals of software engineering, followed by examining an existing framework of a social media application. The students are then grouped into multiple software teams, each focusing on a different aspect of the app. The separate teams must define requirements, design, and provide documentation on the services. Using an Agile development approach, the teams incrementally add to the code base and demonstrate features as the application evolves. Subsequent iterations of the class pick up the prior students' code base, providing experience working with a legacy code base.

Preliminary results of using this approach at the university are presented in this paper including quantitative analysis. Analysis of student software submissions to the cloud-based code repository shows student engagement and contributions over the span of the course. Positive student evaluations show the effectiveness of applying the principles of software engineering to the development of a complex solution in a team environment.

## 1 INTRODUCTION

Students attending traditional undergraduate software engineering classes listen to lectures on software development models and may explore writing critical documents, such as requirements, design, and testing. Compare this to the student experience post-graduation, where the student joins a development team to build software at a company. In the software engineering course the students rarely collectively apply concepts as a member of a team building an application. The student may be surprised by the chasm between the classroom experience and the workplace [1].

This paper investigates combining the study of software engineering principles with the creation of a single software application by the class. Some software engineering classes use the approach of grouping a class into mini-teams of three or four students with each team developing a separate application. This partially addresses criticism by software companies that students are lacking experience in testing, designing, and working in teams. Although the mini-team approach to building an application touches on these areas, it falls short of what the new software engineering graduate will encounter in practice where they may be a member of a large development team [2].

This study reimagines the software engineering class as combining lectures with the building of a single software application with a large team of student developers. To investigate the effectiveness of this approach, a four-month undergraduate software engineering class was restructured to focus on a combination of software engineering concepts and the implementation of a social media application. This paper describes the class design, the observed benefits and drawbacks, and summarizes a quantitative student survey.

## 2 METHODOLOGY

The setting for this study is an undergraduate class composed of 16 junior and senior computer science students taking a second-level software engineering class. The students were required to take the prerequisite first-level class covering software engineering fundamentals. This first-level class used a traditional approach of lectures with minimal small projects. In the restructured second-level class, all students became software developers focused on building a new web social media application.

### 2.1.1 One Code Base, Sixteen Students

Software engineering classes rarely include large-team software development [4]. Some classes may use small teams (six-person or fewer). The advantage of using small fully-independent teams over a single team is less complexity: the application being developed is simpler, there are fewer paths of communication between the members, and one team cannot step on the toes of another. However, the drawback to the small team approach is that it does not fully embody the software engineering challenges in the workplace faced by larger development teams.

Our challenge was to design a course in which small groups (subteams) can work independently but whose work integrates into a single application. Our solution was to use a common architectural approach: create independent service components, one per subteam, that can be assembled into a single application. Knowing that our class would have sixteen students, we carefully selected an application topic divisible into at least six independent services. We chose to develop a simple social media application. This application can be divided into services and has enough complexity for rich interactions between subteams.

Prior to the beginning of the course, we design the architecture for the application and selected MERN (MongoDB, Express, React, Node) as the framework. Fig. 1. shows the service-based architecture. Each service provides a RESTful API for creating, reading, updating, and deleting domain data. These services are the initial focus of student development after creating requirements and designs. The application is then pulled together by the user interface making calls to the services.

Before the class first met, we implemented a skeletal base for the software and created a version control repository. At the beginning of the course, the students selected subteams of two to three developers and chose the service to implement. The student subteams were responsible for developing the selected service and the portion of the user interface primarily relying on the service.

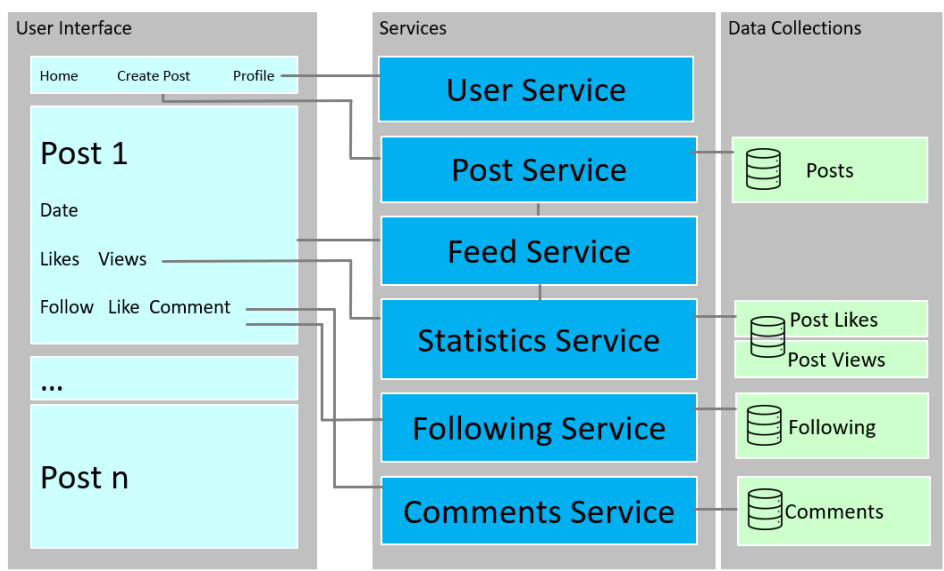

*Figure 1. The service-based architecture of the application*

We selected Agile as the primary software development model but began with a waterfall approach to lay the groundwork. Early lectures discussed requirement development and presented fundamental concepts such as meeting with stakeholders, storyboarding, and creating verifiable requirements. The primary stakeholder was the course instructor in the role of customer. The other service subteams were also stakeholders due to the interconnections between services. For example, the Feed Service requires an API from the Post Service for available posts, and the Following Service for information about whom a user follows to be able to create a compelling feed. Based on stakeholder input, the students created the requirements for their subteam's service and user interface.

Once the requirements were explored we presented software engineering design techniques. The students used these ideas to create service designs. At this point, we transitioned to Agile, created a prioritized backlog of stories, and began implementation with code submissions to the repository (Fig 2.). We held two-week sprints with the instructor as project owner and the scrum master role rotationally filled by students. At each sprint, the subteams demonstrated completed stories by executing code from the repository.

The goal of our software development was to learn the concepts of software engineering and not to complete a fully functional and tested social media app. The time for implementation was relatively short (two months) and the students were not experienced in the technologies being used. The course culminated in a final presentation of the integrated social media application, with each group demonstrating the completed functionality. The students were anonymously and quantitatively surveyed about the course.

## 3 RESULTS

Before the course started, we wondered if this approach to teaching software engineering would be successful. Can a class of students inexperienced with professional software development and the chosen software languages learn software engineering techniques and build a single application in three months? Surprisingly, the answer was yes.

### 3.1 The social media application

The students created a social media application implementing many of the features specified by the requirements. Table 1. lists the completed features. However, the features have a rudimentary user interface and limited error handling.

*Table 1. Completed Features*

| Feature | Description |
| --- | --- |
| User Authentication | User authentication and user sessions. |
| Posting | Creating, viewing, editing, and deleting posts. |
| Private User Profile | Viewing and editing a logged-in user's profile. |
| Public User Profile | Viewing of one user's profile by another. |
| Following | A user following or unfollowing another user. |
| Liking | A user liking or unliking a post. |
| Post Feed | A list of posts prioritized by the users that are followed and the popularity of a post. |

### 3.2 The code, version control, and testing

The student-created code base was approximately 5000 lines of JavaScript code. The code base can be viewed at https://github.com/brockenbrough/social. The students had over 680 commits to the repository during the semester which is approximately forty-three commits each (Fig. 2.). Version control was one of the greatest challenges in the course and the educational topic that benefited the most from large team development. Often during development one student's submitted changes were deleted by another until the students became experienced with the submission process. This was caused by students failing to merge code properly when conflicts occurred. This drove home the importance of properly submitting code and provided a preview of what could happen at the workplace.

Occasionally, a student's code submission would result in breaking the application, blocking other students from developing. When this occurred, the student was identified and strongly encouraged by the other students to resolve the issue. This often occurred due to a lack of proper testing before submission.

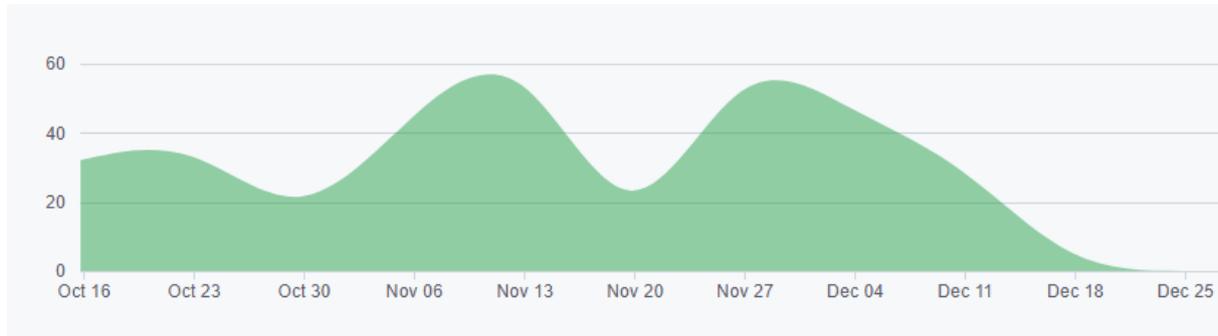

*Figure 2. The number of student contributions by day to the code base during the semester.*

### 3.3 Results from the survey of software engineering students

The students in the class were surveyed on their reactions to learning software engineering through a large team-based approach.

*3.3.1 Software engineering topics benefiting from this approach*

Fig. 3. shows a ranking of the software engineering topics that benefited from this approach to learning software engineering skills. Understanding version control and software management were by far viewed as benefitting the most from team software development. Version control is essential to coordinating the development work of multiple people. Without proper understanding and use of a version control system submitted code can be lost as multiple developers attempt to make changes to the same file. Few students in the class had used version control before on a multiple-person project. Prior experience was limited to single-person or two-person projects where a collision of changes is unlikely.

During development, there were many instances of one person's submitted change being overwritten by another due to improper merging of the changed files. The lecture portion of the class described how a distributed version control system functioned, but only through practice do the issues become apparent. The interactions between students who lost code and those who wiped out others' changes were sometimes heated. By the end of development, code was no longer being lost and all understood the importance of applying version control properly.

Design and team communication were ranked highest as benefitting from the team-based approach. During design, there were many cross-subteam discussions of how services should be structured and how the API results should be provided. Not surprisingly, understanding team communication also benefitted from discussions on requirements, design, implementation, and version control issues.

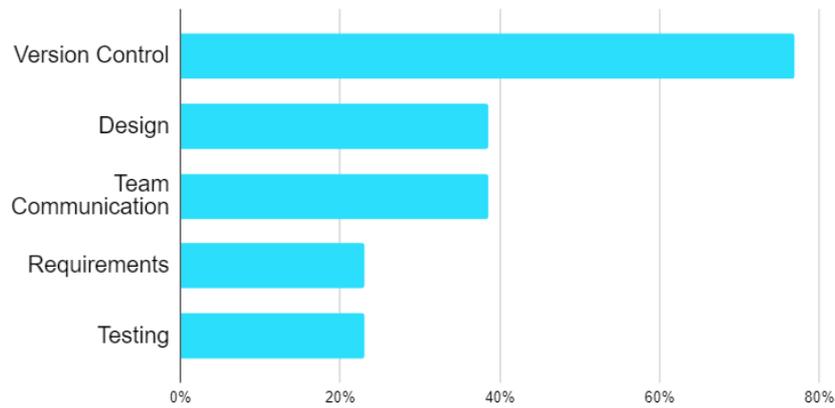

*Figure 3. Results from a survey of students of the importance of the team-based application development approach in the learning of software engineering skills.*

### 3.3.2   Ideal team size

In other undergraduate computer science courses at our university, the typical team size is two to four people. As the team grows in size, communication between developers becomes more complex and coordinating changes more difficult. The students were given three choices for the ideal team size for learning software engineering. The choices were (1) 2 to 4 people, (2) 5 to 6 people, or (3) 10 to 20 people. In considering this choice the students were faced with opposing views: smaller projects are easier to complete but larger projects better demonstrate the software engineering issues. As shown in Fig 4. the top selection for the ideal team size for learning was 10 students or more, and the majority of the students chose an ideal team size of 5 people or higher.

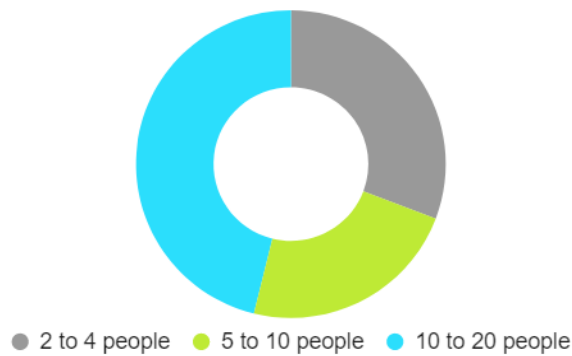

*Figure 4. Student survey of ideal team size for learning software engineering.*

### 3.3.3   Usefulness of team development in understanding concepts

The students were asked about a variety of software engineering topics to indicate to what degree team development was important in gaining an understanding of the concept. Fig 5. shows the results of this survey question. Team development was once again viewed as very important to understanding version control and teamwork. Team development was viewed as slightly less important for understanding Agile. This may be due to the difficulty of running an Agile project in the classroom when only meeting two days a week.

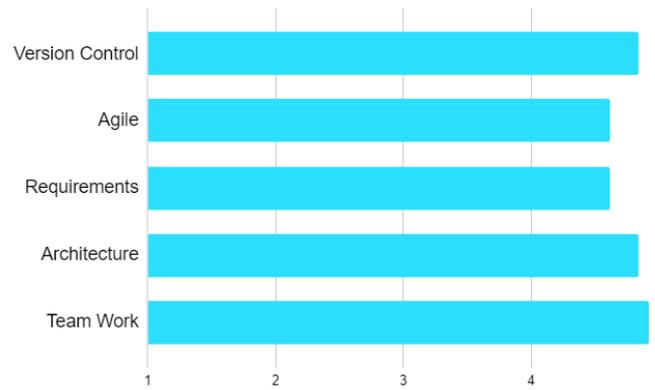

Figure 5. Student survey the importance of team development "not that important" (1) to "very important" (5).

## 4 CONCLUSIONS

Creating a software engineering course that combines traditional lectures with the student-based development of a single application is challenging: an application must be carefully selected that provides independence between components, the students must be well-versed in version control prior to submitting changes, and the instructor must be prepared to guide the students to resolve blocking issues.  However, we have shown that there are great benefits to using this approach.  The students gain a deeper understanding of software engineering topics.  The concept of version control comes to life as students work from a single code base and experience merge collisions that must be resolved.  The complexity of team communication is explored as students must understand the needs of other student developers and provide working solutions.  The large-team approach culminates in the collective pride of the students as their software engineering work comes together in a finished application.

The next offering of this course will build upon this initial class offering.  The new students will use the prior classes' code base as the starting point and must learn how to work with legacy code.  They will learn how to understand a large working code base and make improvements without breaking the functionality.  This will add another dimension to this approach by providing an experience similar to what new graduates will experience as they enter the software development workforce [2].